# The electronic structure and intervalley coupling of artificial and genuine graphene superlattices

Pilkwang Kim[1] and Cheol-Hwan Park[1*]

[1]*Department of Physics and Astronomy, Seoul National University, Seoul 08826, Korea*

## Abstract

A so-called artificial graphene is an artificial material whose low-energy carriers are described by the massless Dirac equation. Applying a periodic potential with triangular symmetry to a two-dimensional electron gas is one way to make such a material. According to recent experimental results, it is now possible to realize an artificial graphene in the lab and to even apply an additional lateral, one-dimensional periodic potential to it. We name the latter system an *artificial graphene superlattice* in order to distinguish it from a genuine graphene superlattice made from graphene. In this study, we investigate the electronic structure of artificial graphene superlattices, which exhibit the emergence of energy band gaps, merging and splitting of the Dirac points, etc. Then, from a similar investigation on genuine graphene superlattices, we show that many of these features originate from the coupling between Dirac fermions residing in two different valleys, the intervalley coupling. Furthermore, contrary to previous studies, we find that the effects of intervalley coupling on the electronic structure cannot be ignored no matter how long the spatial period of the superlattice is.

[*] Address correspondence to Cheol-Hwan Park, cheolhwan@snu.ac.kr

# 1. Introduction

Tuning the electronic structure of graphene has been one of the most popular research subjects since its experimental isolation. For example, various techniques of introducing dopants into graphene in order to control the density of states and chemical potential have been developed [1-4]. Also, if graphene is strained, the electrons feel valley-dependent pseudo magnetic field [5-7]. Applying an additional periodic potential is another way to modulate the electronic structure [8]; recently, the fractal quantum Hall effect in graphene on hexagonal boron nitride, or, a graphene Moiré superlattice, was observed [9-12].

Contrary to the methods of using external perturbations, completely different approaches, making an artificial material whose electronic structure is similar to that of natural graphene, have been proposed [13-19]. This so-called artificial graphene has vanishing energy gaps at two points in the Brillouin zone (the Dirac points) around which the energy-momentum dispersion is linear, similarly to graphene. An artificial graphene can be made by applying a periodic potential with triangular symmetry to a two-dimensional electron gas (2DEG) [13, 14]. Similar proposals for photonic crystals [20, 21] and for ultracold atomic gases [22, 23] have been reported. Electronic and optical signatures of massless Dirac fermions have been experimentally observed in various types of artificial graphene systems [14-19, 23]. The artificial graphene has been actively studied recently [24-39].

For an obvious reason, there are much more possibilities in tuning the electronic structure of artificial graphene than genuine graphene. The spatial period, amplitude and overall profile of the applied potential as well as the strength of the atomic spin-orbit coupling can largely be tuned.

By taking advantage of these additional degrees of freedom, one can study emergent phenomena in artificial graphene which were not accessible in natural graphene. The Mott-Hubbard transition and other phenomena originating from electron-electron interactions were reported [15, 40, 41]. Also, a quantum spin Hall phase was predicted [28, 31].

The recently reported molecular artificial graphenes are made either by collecting small molecules (such as CO and coronene) with the tip of a scanning tunneling microscope or by self-assembling supramolecules. These molecules form a triangular array on a metallic surface as schematically shown in Fig. 1(a), upper side. Then, the metallic surface states feel the local potential induced by the adsorbed molecules. PN junctions, Kekulé-type distortions, strains inducing a quantum Hall effect without an actual magnetic field, and several types of artificial

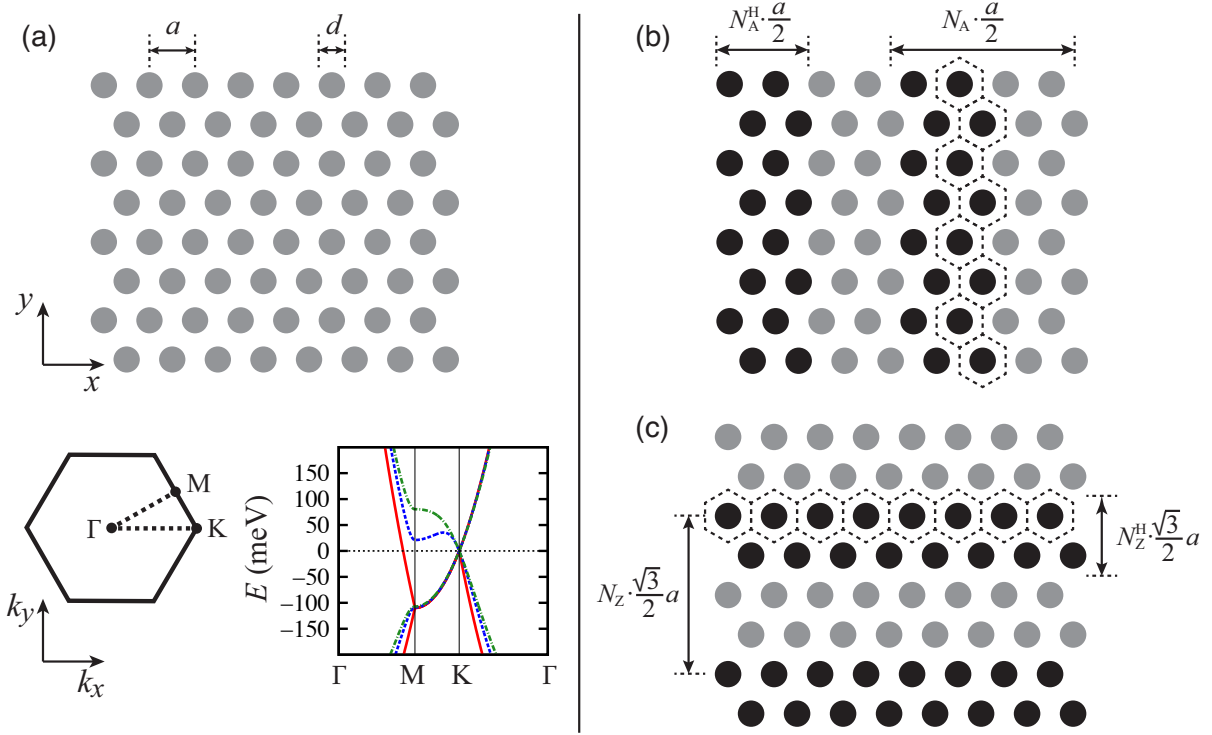

**Figure 1** (a) A schematic of artificial graphene made from a muffin-tin potential (upper side). The potential is $V_T$ inside the disks and zero outside. The calculated energy bandstructures for artificial graphene along the path in the Brillouin zone (bottom left side) are shown in the bottom right side; solid or red, dashed or blue, and dash-dotted or green curves show the results for $V_T = 0.0$, 0.5, and 1.0 eV, respectively. We have used $a = 20$ Å, $d = 13.3$ Å, and $m^* = 0.38 m_0$. The Dirac point energy at K is set to zero. (b) A schematic of an armchair-type artificial graphene superlattice. Darker disks represent regions with higher potential. In experiments, the potential difference between disks can be realized (i) by using two different kinds of molecules, (ii) by tuning the number of the same molecules located at each lattice site, or (iii) by changing local electrostatic potential with a gate electrode. (c) A schematic similar to (b) for a zigzag-type artificial graphene superlattice.

defects were implemented in the molecular artificial graphenes and many theoretical predictions were confirmed [17-19].

Since there are several kinds of molecules that can be used to make an artificial graphene, one can think of a superlattice of artificial graphene made with different types of molecules or different numbers of molecules per lattice site or a one-dimensional (1D) superlattice potential generated by an external gate (Fig. 1(b)).

In the former part of this paper, we examine the electronic structure of artificial graphene superlattices which are made by applying additional 1D periodic potentials to artificial graphenes. Previous studies on graphene under such a potential, which we call a genuine graphene superlattice in order to distinguish it from an artificial graphene superlattice, have reported anisotropic group velocity renormalization, emergence of new zero-energy points, conductance resonance, etc [8, 42-48]. Artificial graphene superlattices studied in this paper exhibit further features including the displacement of the Dirac points in momentum space and merging and splitting of the Dirac points which are not previously reported.

To find out why those features were not anticipated in the previous studies, we revisit the electronic structures of graphene superlattices in the latter part of this paper. We find that such characteristic features originate from the intervalley coupling.

Furthermore, we demonstrate that the effects of the intervalley coupling should not be neglected even if the spatial period of the superlattice potential is long (i. e., even if the reciprocal lattice vector of the superlattice potential is much shorter than the distance between the two Dirac points). This seemingly counterintuitive finding is in disagreement with the underlying assumption of the former studies (Ref. [8] and related theoretical studies) that the intervalley coupling can be readily ignored in superlattices with spatial periods that are long enough.

## 2. Artificial graphene superlattices

### 2.1. Nearly-free-electron model

An artificial graphene can be made by putting a 2DEG (either a conventional 2DEG or a metallic surface state) under the effect of a muffin-tin-like repulsive potential with triangular symmetry (Fig. 1(a)) [13]. As the potential strength is increased, the energy window in which the energy-momentum dispersion relation is linear becomes wider (Fig. 1(a) bottom right).

An artificial graphene superlattice is made by applying an additional 1D periodic potential to this artificial graphene. Here, we consider two different types of artificial graphene superlattices: armchair-type and zigzag-type as shown in Figs. 1(b) and 1(c).

The Hamiltonian describing the charge carriers in an artificial graphene superlattice reads

$$H = H_0 + V_1 + V_2 \qquad (1)$$

where $H_0 = -\hbar^2 \nabla^2/2m^*$ is the kinetic energy operator and $m^*$ the effective mass of the carriers. We set $m^* = 0.38\, m_0$, where $m_0$ is the mass of an electron, to model the surface states of Cu (111) used in the experiments on molecular artificial graphene [17-19]. We denote the muffin-tin-type potential with triangular symmetry by $V_1 = \sum_\tau V_0(\mathbf{r} - \tau)$ where $\tau$'s are the lattice vectors of a triangular lattice whose lattice parameter is $a$. Here,

$$V_{\text{o}}(\mathbf{r}) = \begin{cases} V_{\text{T}}, & |\mathbf{r}| \leq d/2 \\ 0, & \text{otherwise,} \end{cases} \tag{2}$$

where $V_{\text{T}}$ ($> 0$) is the strength of the muffin-tin-type potential. We set $a = 2\text{nm}$, $d = 1.32\text{nm}$, and $V_{\text{T}} = 1.0eV$. (We choose $d$ to maximize the relevant Fourier components of the muffin-tin-type potential [13].)

An artificial graphene (Fig. 1(a)) is described by the Hamiltonian $H_0 + V_1$. An eigenstate of this Hamiltonian with Bloch wavevector $\mathbf{K} + \mathbf{k}$, where $\mathbf{K} = (4\pi/3a, 0)$, $|\mathbf{K} + \mathbf{k}\rangle$ can be approximately written as

$$\langle \mathbf{r}|\mathbf{K} + \mathbf{k}\rangle \approx \frac{1}{\sqrt{3A_{\text{c}}}}\left[c_1 e^{i(\mathbf{K}_1+\mathbf{k})\cdot\mathbf{r}} + c_2 e^{i(\mathbf{K}_1+\mathbf{k})\cdot\mathbf{r}} + c_3 e^{i(\mathbf{K}_1+\mathbf{k})\cdot\mathbf{r}}\right] \tag{3}$$

where $A_{\text{c}}$ is the total area of the crystal and $\mathbf{K}_1 = (4\pi/3a, 0)$, $\mathbf{K}_2 = (-2\pi/3a, 2\pi/\sqrt{3}a)$, and $\mathbf{K}_3 = (-2\pi/3a, -2\pi/\sqrt{3}a)$ are the corners of the first Brillouin zone interconnected by the reciprocal lattice vectors. The wavefunction $\psi_{\mathbf{K}+\mathbf{k}}(\mathbf{r})$ can simply be represented by $\mathbf{c} = (c_1, c_2, c_3)^{\text{T}}$. Two of the three lowest-energy eigenstates of $H_0 + V_1$ are degenerate at $\mathbf{K}$. The two eigenstates with momentum $\mathbf{K} + \mathbf{k}$ associated with the degenerate states at $\mathbf{K}$ are

$$\mathbf{c}(1, \mathbf{k}) = \sqrt{\frac{2}{3}}\left[\cos\frac{\theta_{\mathbf{k}}}{2}, \cos\left(\frac{\theta_{\mathbf{k}}}{2} + \frac{2\pi}{3}\right), \cos\left(\frac{\theta_{\mathbf{k}}}{2} + \frac{4\pi}{3}\right)\right]^{\text{T}}$$

$$\mathbf{c}(-1, \mathbf{k}) = \sqrt{\frac{2}{3}}\left[\sin\frac{\theta_{\mathbf{k}}}{2}, \sin\left(\frac{\theta_{\mathbf{k}}}{2} + \frac{2\pi}{3}\right), \sin\left(\frac{\theta_{\mathbf{k}}}{2} + \frac{4\pi}{3}\right)\right]^{\text{T}}, \tag{4}$$

where $\theta_{\mathbf{k}} = \tan^{-1}(k_y/k_x)$ [13]. The eigenvalues corresponding to $\mathbf{c}(1, \mathbf{k})$ and $\mathbf{c}(-1, \mathbf{k})$ are $\hbar v_0 \mathbf{k}$ and $-\hbar v_0 \mathbf{k}$, respectively, where $v_0 = 2\pi\hbar/3m^*a$. The eigenstates whose Bloch wavevector is around $\mathbf{K}'$ can similarly be found [13].

$V_2$ in Eq. (1) describes an additional 1D periodic potential:

$$V_2(x) = \sum_n V_{\text{1D,A}}\left(x - \frac{a}{2}N_A n\right), \tag{5}$$

in the case of a zigzag-type superlattice. Here, $N_A$ and $N_Z$ is the number of unit cells of the original artificial graphene contained in a unit cell of the artificial graphene superlattice (see Figs. 1(b) and 1(c));

$$V_{1D,A}(x) = \begin{cases} V_{SL}, & -a/4 < x \leq N_A^H a/2 - a/4 \\ 0, & \text{otherwise,} \end{cases} \quad (6)$$

and

$$V_{1D,Z}(y) = \begin{cases} V_{SL}, & -\sqrt{3}a/4 < y \leq \sqrt{3}N_Z^H a/2 - \sqrt{3}a/4 \\ 0, & \text{otherwise,} \end{cases} \quad (7)$$

where $N_A^H$ or $N_Z^H$ is the number of unit cells with higher potential (Figs. 1(b) and 1(c)). The primitive vectors of the lattice of an armchair-type superlattice are

$$\mathbf{a}_1 = \left(\frac{N_A a}{2}, \frac{\sqrt{3}N_A a}{2}\right) \text{ and } \mathbf{a}_2 = (0, \sqrt{3}a)$$

and the corresponding primitive vectors of the reciprocal lattice are

$$\mathbf{b}_1 = \left(\frac{4\pi}{N_A a}, 0\right) \text{ and } \mathbf{b}_2 = \left(-\frac{2\pi}{a}, \frac{2\pi}{\sqrt{3}a}\right). \quad (8)$$

The matrix element

$$\begin{aligned}
\langle \mathbf{k}'|V_1|\mathbf{k}\rangle &= \sum_{\mathbf{G}} \delta_{\mathbf{k}',\mathbf{k}+\mathbf{G}} V_1(\mathbf{G}) \\
&= \sum_{\mathbf{G}} \delta_{\mathbf{k}',\mathbf{k}+\mathbf{G}} \frac{1}{N_A} \frac{2}{\sqrt{3}a^2} \sum_{\tau} \int e^{-i\mathbf{G}\cdot\mathbf{r}} V_o(\mathbf{r}-\tau) d\mathbf{r} \\
&= \sum_{\mathbf{G}} \frac{2}{\sqrt{3}a^2} \delta_{\mathbf{k}',\mathbf{k}+\mathbf{G}} S(\mathbf{G}) f(\mathbf{G})
\end{aligned}$$

is the scattering amplitude between the two plane-wave states $|\mathbf{k}\rangle$ and $|\mathbf{k}'\rangle$. Here, $\mathbf{G} = l_1 \mathbf{b}_1 + l_2 \mathbf{b}_2$ is a reciprocal lattice vector,

$$S(\mathbf{G}) = S(l_1 \mathbf{b}_1 + l_2 \mathbf{b}_2)$$

$$= \frac{1}{N_A} \sum_{m=0}^{N_A-1} \exp\left(-i l_1 \frac{4\pi}{N_A a} \frac{a}{2} m\right)$$

$$= \begin{cases} 0, & l_1 \neq N_A l \\ 1, & l_1 = N_A l \end{cases}$$

is the structure factor for an integer $l$ and, from Eq. (2),

$$f(\mathbf{G}) = \int e^{-i\mathbf{G}\cdot\mathbf{r}} V_o(\mathbf{r}) d\mathbf{r}$$

$$= \frac{2\pi d V_T J_1(|\mathbf{G}|d)}{|\mathbf{G}|},$$

where $J_1(x)$ is the Bessel function of the first kind, is the atomic form factor of $V_o$. On the other hand,

$$\langle \mathbf{k}' | V_2 | \mathbf{k} \rangle = \sum_{\mathbf{G}} \delta_{\mathbf{k}', \mathbf{k}+\mathbf{G}} V_2(\mathbf{G})$$

$$= \sum_{l_1} \delta_{k'_x, k_x + l_1 \frac{4\pi}{N_A a}} \delta_{k'_y, k_y} \frac{2}{N_A a} \int_{-N_A a/4}^{N_A a/4} e^{-i(k_x - k'_x)x} V_2(x)$$

$$= \sum_{l_1} \frac{V_{SL}}{\pi l_1} e^{i\pi l_1 (N_A^H - 1)/N_A} \sin\left(\frac{N_A^H}{N_A} \pi l_1\right) \delta_{k'_x, k_x + l_1 \frac{4\pi}{N_A a}} \delta_{k'_y, k_y}$$

(9)

We have used Eq. (5) in the last equality. The corresponding matrix elements for a zigzag-type superlattice can be derived in a similar way.

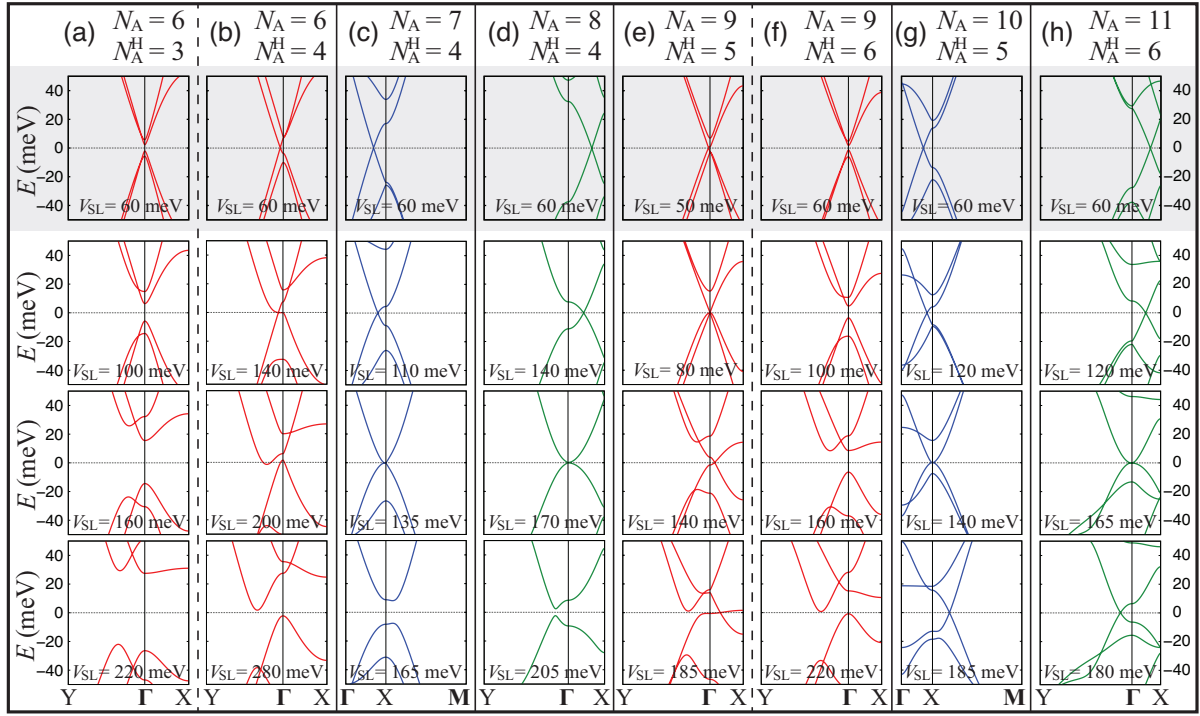

**Figure 2** The calculated energy band structures of armchair-type artificial graphene superlattices. We have used $V_T = 1.0$ eV, $a = 20$ Å, $d = 13.3$ Å, and $m^* = 0.38\, m_0$. $\Gamma = (0,0)$, X $= (2/N_A, 0)$, Y $= (0, 1/\sqrt{3})$, and M $= (2/N_A, 1/\sqrt{3})$ in units of $2\pi/a$. Dirac point energy is set to zero.

Previous studies on graphene superlattices, which assumed only one Dirac cone, have proposed that new zero-energy modes (or, Dirac cones) are generated around the original Dirac point if the amplitude of the additional 1D periodic potential is larger than a certain value [45, 46]. We define $V_c$ as the minimum $V_{SL}$ (see Eqs. (6) and (7)) generating such new zero-energy modes, which is given by [45]

$$V_c = \begin{cases} \dfrac{4\pi\hbar v_0}{N_A \cdot a/2}, & \text{an armchair– type superlattice} \\ \dfrac{4\pi\hbar v_0}{N_Z \cdot \sqrt{3}a/2}, & \text{a zigzag– type superlattice.} \end{cases} \quad (10)$$

## 2.2. Armchair-type superlattices with $N_A = 3n$

In this subsection, we discuss armchair-type artificial graphene superlattices with $N_A$ being a multiple of three. The Dirac points located at K and K′ in the Brillouin zone of an artificial graphene are folded into $\Gamma$ in the Brillouin zone of the superlattice. The four eigenstates (two from K and the other two from K′) are degenerate in energy if $V_{SL} = 0$; this degeneracy at Γ is lifted if $V_{SL} \neq 0$.

In this case, whether $N_A^H$ is divisible by three or not largely determines the electronic structure. If $N_A^H$ is also a multiple of three, the energy-momentum dispersion is quadratic and an energy window in which the density of states vanishes appears (top-most, shaded panels in Figs. 2(a) and 2(f) and Fig. 3(b)).

If $N_A^H$ is not divisible by three, such a zero-density-of-states energy window does not exist (top-most, shaded panels in Figs. 2(b) and 2(e) and Fig. 3(c)). The Dirac point initially located at Γ splits into two different zero-energy points; the two points move along Γ − Y which is perpendicular to the periodic direction of the superlattice potential (Fig. 3(c)). The density of states varies linearly with energy in the low-energy regime (Fig. 3(c)), similarly to that of an artificial graphene without superlattice potential (Fig. 3(a)). Van Hove singularities in the density of states (e. g., near −4 meV and 7 meV on the right of Fig. 3(c)) are attributed to the saddle points formed at the Γ point (left side of Fig. 3(c)).

We can obtain the approximate energy splitting at Γ originating from the effects of intervalley coupling by using perturbation theory. Eigenstates which were originally located at K are given by Eq. (4) and those from K′ are the time reversal pairs of Eq. (4). The energy shifts at Γ from first-order degenerate perturbation theory are

$$\Delta E = \pm |A|, \pm \frac{1}{3}|A + 2B|, \qquad (11)$$

where

$$\begin{aligned} A &= \frac{2}{N_A a} \int_{-N_A a/4}^{N_A a/4} e^{i\frac{4\pi}{3a}} V_2(x) dx \\ &= V_{SL} e^{i\frac{\pi}{3}(N_A^H - 1)} \frac{3}{\pi N_A} \sin\left(\frac{\pi}{3} N_A^H\right) \\ B &= \frac{2}{N_A a} \int_{-N_A a/4}^{N_A a/4} e^{i\frac{8\pi}{3a}} V_2(x) dx \\ &= V_{SL} e^{i\frac{2\pi}{3}(N_A^H - 1)} \frac{3}{2\pi N_A} \sin\left(\frac{2\pi}{3} N_A^H\right) \end{aligned} \qquad (12)$$

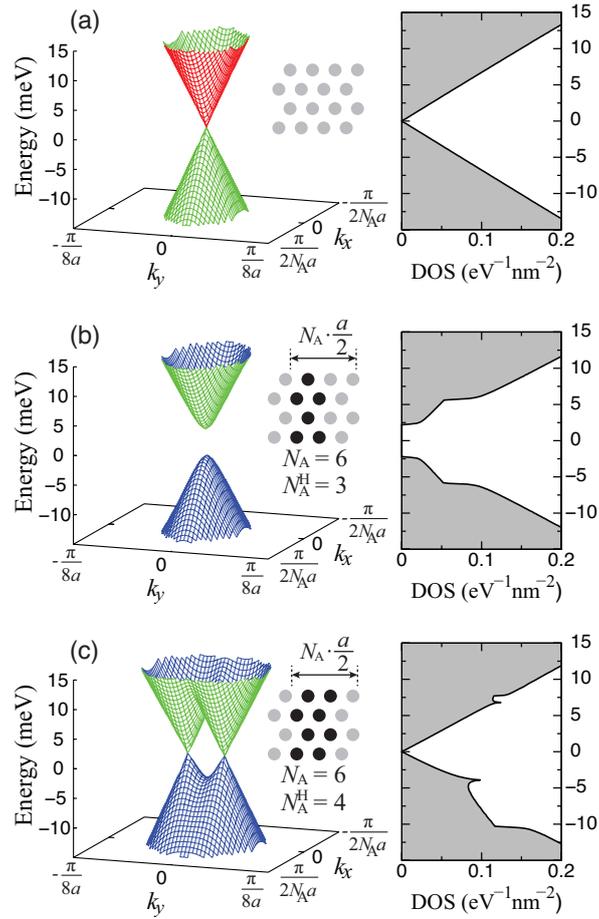

**Figure 3** (a) The energy band structure around Γ and density of states of armchair-type artificial graphene superlattice with $N_A = 6$ when no additional one-dimensional potential is applied. (b) and (c) Similar quantities as in (a) but with $N_A^H = 3$ and $N_A^H = 4$, respectively. We have used $V_T = 1.0$ eV and $V_{SL} = 60$ meV.

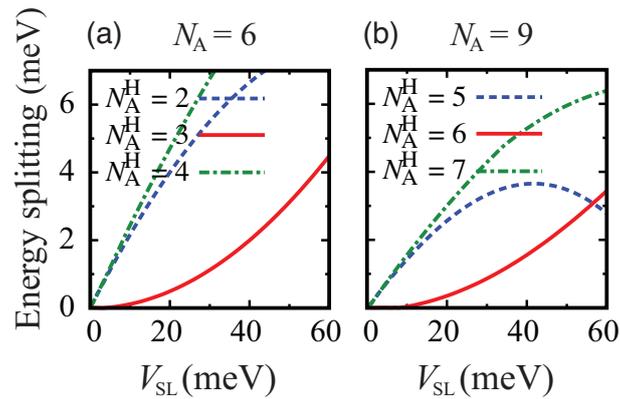

**Figure 4** The energy splitting between the second and third energy eigenvalues at the Dirac point (Γ in our case) for armchair-type artificial graphene superlattices. If $V_{SL} = 0$, the four eigenstates are all degenerate.

We have used Eqs. (5) and (6) for $V_2(x)$. Note that $A$ and $B$ remain finite if $N_A^H$ is not a multiple of three while they vanish if $N_A^H$ is. Therefore, when $V_{SL}$ is low, the size of the energy splitting grows linearly with $V_{SL}$ if $N_A^H$ is not a multiple of three, whereas it grows quadratically with $V_{SL}$ if $N_A^H$ is a multiple of three due to the absence of first-order correction to the energy splitting (Fig. 4).

We find that the energy splitting is inversely proportional to $N_A$. If $N_A^H$ is not a multiple of three, the energy splitting at $\Gamma$ (Eq. (11)) are inversely proportional to $N_A$ (Eq. (12)). If it is, the energy shift from second-order perturbation theory is,

$$\sim \frac{|\langle\langle \mathbf{K}' \pm 4\pi/N_A a|V_2|\mathbf{K}\rangle\rangle|^2}{4\pi\hbar v_0/N_A a} \propto \frac{1/N_A^2}{1/N_A} = \frac{1}{N_A}, \qquad (13)$$

where $|\mathbf{K}\rangle\rangle$ and $|\mathbf{K}' \pm 4\pi/N_A a\rangle\rangle$ are the eigenstates (Eq. (3)) of $H_0 + V_1$ (see Eq. (1)). (Note that the matrix elements of $V_2$ between planewave states (Eq. (9)) are inversely proportional to $N_A$. Since $|\mathbf{K}\rangle\rangle$ and $|\mathbf{K}' \pm 4\pi/N_A a\rangle\rangle$ are linear combinations of the planewave states, $|\langle\langle \mathbf{K}' \pm 4\pi/N_A a|V_2|\mathbf{K}\rangle\rangle|$ is also inversely proportional to $N_A$.)

On the other hand, the energy bandwidth is also inversely proportional to the spatial period because the energy-momentum dispersion is linear in the low-energy regime. Thus, the effects of intervalley coupling on the electronic structure cannot be ignored no matter how long the spatial period is.

One may think that intervalley coupling has a negligible effect on the electronic structure if the length of the reciprocal lattice vector of the superlattice is much shorter than the distance between K and K' in momentum space, i.e., if the spatial period of the superlattice is sufficiently long. However, we have found that a long spatial period does not guarantee that the effects of intervalley coupling in artificial graphene superlattices are negligible.

### 2.3. Armchair-type superlattices with $N_A = 3n \pm 1$

We define *single-valley approximation*, as the approximation neglecting the effect of intervalley coupling. According to previous studies on graphene superlattices adopting this scheme, (i) the group velocity perpendicular to the periodic direction decreases with $V_{SL}$. (ii) When $V_{SL}$ is equal to $V_c$ (Eq. (10)), the group velocity vanishes [8, 43, 49] and the energy-momentum dispersion becomes quadratic along that direction. (iii) If $V_{SL}$ is larger than $V_c$,

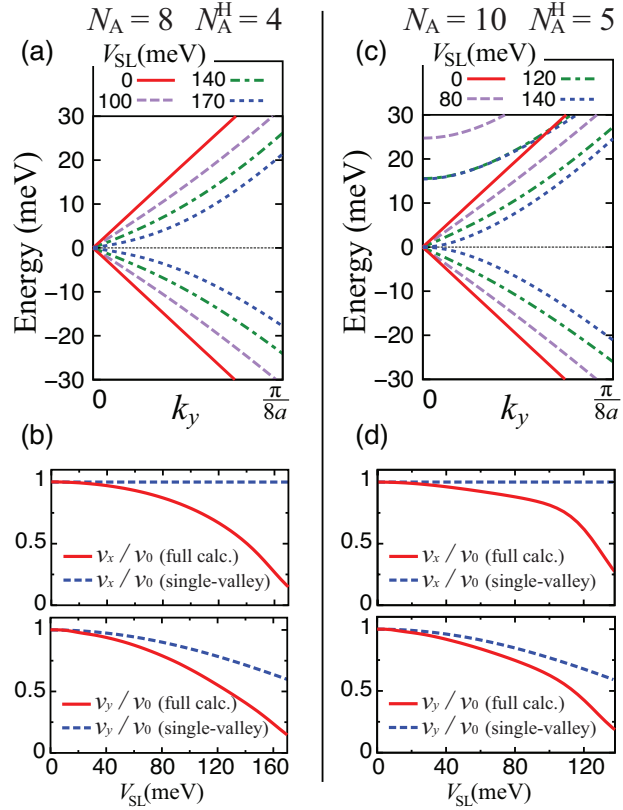

**Figure 5** (a) and (c) The energy-momentum dispersions along the $k_y$ direction of armchair-type artificial graphene superlattices. $k_y = 0$ in each curve corresponds to the Dirac point (whose position in the Brillouin zone varies with $V_{SL}$). (b) and (d) The group velocities at the Dirac point divided by $v_0$, the value at $V_{SL} = 0$. Solid or red curves are the results of the full calculations; dashed or blue curves are obtained from a theory neglecting intervalley coupling effects.

two additional Dirac cones are generated and move along the $k_y$ direction, which is perpendicular to the periodic direction, but not along the $k_x$ direction [45, 46, 50]. (iv) The group velocity along the periodic direction of the superlattice potential is not reduced [45, 46]. According to our calculations, the group velocity along the $k_y$ direction indeed decreases with $V_{SL}$ (Figs. 5(a) and 5(c)) but much faster than the expectations made with single-valley approximation (lower panels of Figs. 5(b) and 5(d)). In addition, the group velocity along the $k_x$ direction, which is the periodic direction of the superlattice potential, is also significantly reduced (upper panels of Figs. 5(b) and 5(d)).

Now we focus on the artificial graphene superlattice with $N_A = 10$ and $N_A^H = 5$ (Figs. 2(g), 5(c), and 5(d)). The electronic band structure calculated without the single-valley approximation is qualitatively different from that calculated within the approximation. To start with, the Dirac point moves along the $k_x$ direction (the second row of Fig. 2(g) and Fig 6(a)); in contrast to the previous studies based on single-valley approximation [45, 46, 50].

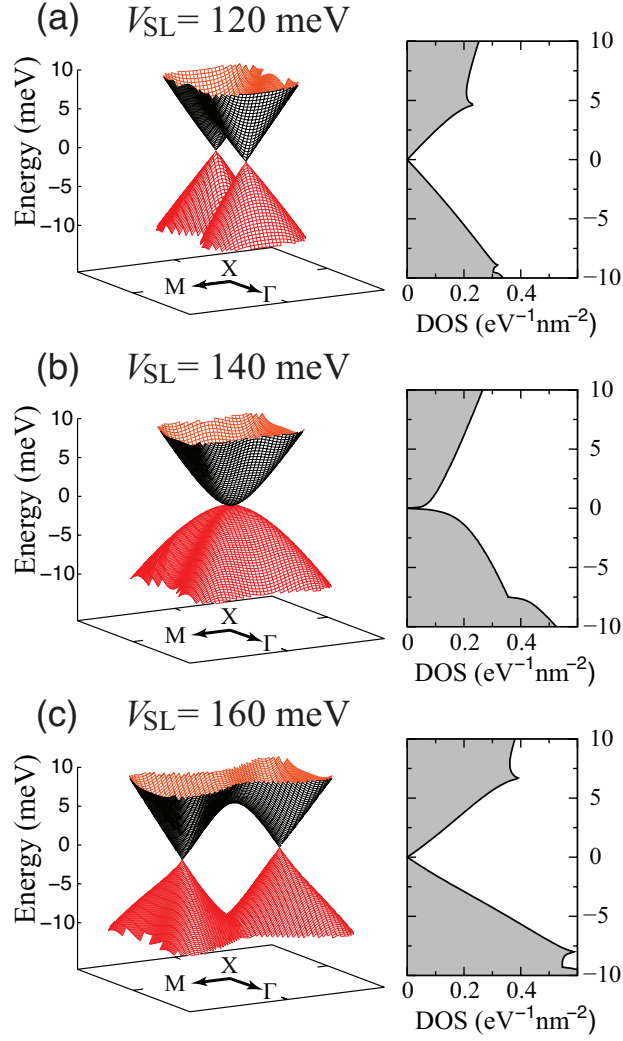

**Figure 6** The energy band structures around X and densities of states of armchair-type artificial graphene superlattices with $N_A = 10$ and $N_A^H = 5$.

As $V_{SL}$ increases, a Dirac point approaches X in the Brillouin zone of the superlattice and merges with another one approaching X from the opposite side, when the energy-momentum dispersion is quadratic along every direction (Fig. 6(b)). If $V_{SL}$ is increased further, the quadratic touching point is split back into two different zero-energy points having linear energy-momentum dispersions around them; however, they move along X − M (Fig. 6(c)) which is perpendicular to the direction along which the two Dirac points move before they merge (upon increasing $V_{SL}$). Ther merging point is X if $N_A = 3n + 1$ (Figs. 2(c) and 2(g)) and Γ if $N_A = 3n - 1$ (Figs. 2(d) and 2(h)). For further analysis, we define $\tilde{V}_c$ as the value of $V_{SL}$ at which the two Dirac points merge (at Γ or X). ($\tilde{V}_c$ is meaningful only if $N_A = 3n \pm 1$).

The energy-momentum dispersion of a superlattice with $V_{SL} = \tilde{V}_c$ around the merging point (Γ or X) falls into one of the two categories: (i) quadratic along $k_x$, the periodic direction, and linear along $k_y$ (Figs. 7(a), 7(b), 7(e), and 7(f)) and (ii) quadratic along both directions (Fig. 6(b) and Figs. 7(c), 7(d), 7(g), and 7(h)). In the former case, the zero energy point disappears and the energy gap opens up as $V_{SL}$ is increased further from $\tilde{V}_c$ (see Figs. 2(c) and 2(d)), while in the latter case, the zero energy points do not disappear but moves along $k_y$ as shown in Fig. 6 (see also Figs. 2(g) and 2(h)). It is noteworthy that the former case corresponds to the type-I semi-Dirac model in Huang *et al* [51].

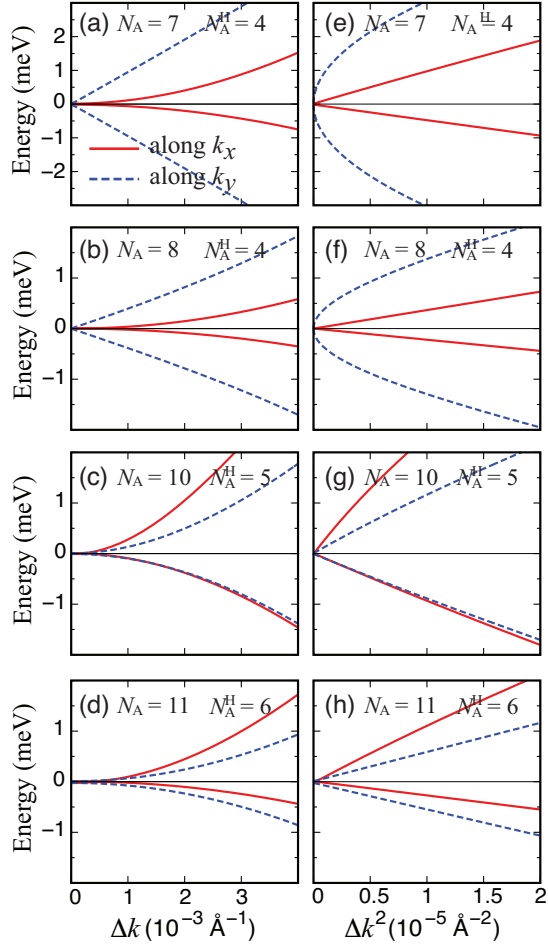

**Figure 7** The energy band structures of artificial graphene superlattices with $V_{SL} = \tilde{V}_c$ (see text) around the merging point of the two Dirac points (Γ or X). Here, $\Delta k$ is the distance in reciprocal space of **k** from the merging point.

If $N_A$ is not a multiple of three, the first order correction from the scattering between states at K and K′ vanishes. However, by extending the argument around Eq. (13) to higher-order perturbations, we find that the energy shifts in our case ($N_A = 3n \pm 1$) arising from intervalley coupling is inversely proportional to $N_A$, or the spatial period of the superlattice. On the other hand, the bandwidth in the Brillouin zone of the superlattice is also inversely proportional to $N_A$. Since both quantities are inversely proportional to $N_A$, the inter-valley coupling effects, again, cannot be neglected in cases $N_A = 3n \pm 1$ even if the spatial period is long.

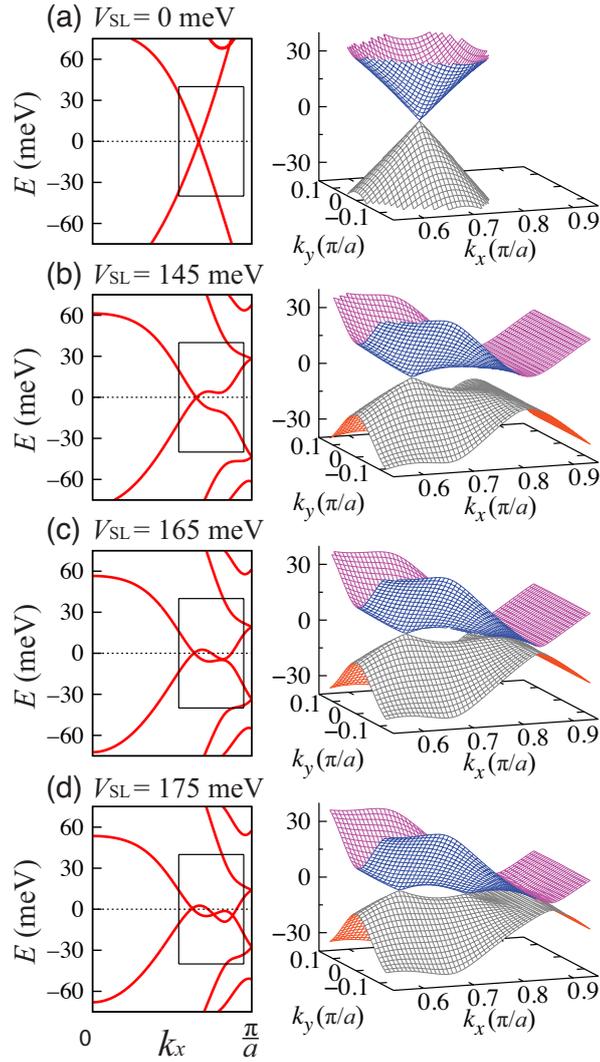

**Figure 8** The energy band structures of zigzag-type artificial graphene superlattices. The box inside the figure on the left hand side illustrates the energy-momentum range to be used in the figure on the right hand side.

## 2.4. Zig-type superlattices

In this section, we investigate the zigzag-type artificial graphene superlattices, in which the intervalley coupling is strongly suppressed. Since the $k_x$ component of the states close to one Dirac point is substantially different from that associated with the other Dirac point, the effects of intervalley coupling involving a superlattice potential periodic along $y$ (Fig. 1(c)) are small. (One cannot say, however, that the effects are absent; the Brillouin zone cannot be divided into two mutually exclusive regions associated with the two Dirac points.)

The group velocity along $k_x$, which is perpendicular to the periodic direction of the zigzag-type superlattice potential, decreases with $V_{SL}$ (Fig. 8(b)). The upper and lower bands on the right side of the Dirac point approach each other until they are degenerate (Fig. 8(c)). Consequently, the two bands intersect each other and two new zero-energy points are created (Fig. 8(d)).

The lifting of the degeneracy (or the opening of a gap) at the Dirac points (Fig 3(b)), displacement of the Dirac point along the periodic direction (Figs. 6(a) and 6(b)), and merging and splitting of the Dirac points (Fig. 6) do not occur in the electronic structure of zigzag-type superlattices; the results support our proposal that such characteristic features originate from intervalley coupling.

## 3. Genuine graphene superlattices

### 3.1. Tight-binding model

In this section we revisit the electronic structure of graphene under a 1D periodic potential, which we call a genuine graphene superlattice in order to distinguish it from an artificial graphene superlattice. We have considered armchair-type superlattices only (Fig. 9) to investigate the effects of intervalley coupling. We have used a tight-binding model containing nearest-neighbor hopping processes between $p_z$ orbitals, whose amplitude $t$ is set to $-3\ eV$. The on-site potential for $i$-th $p_z$ orbital is

$$V_i = \begin{cases} V_{SL}, & l\dfrac{N_A a}{2} - \dfrac{a}{4} < x_i \le l\dfrac{N_A a}{2} + \dfrac{N_A^H a}{2} - \dfrac{a}{4} \\ & l = \ldots -2, -1, 0, 1, 2, \ldots \\ 0, & \text{otherwise,} \end{cases} \quad (14)$$

where $x_i$ is the $x$ coordinate of the $i$-th $p_z$ orbital. $N_A$ is the total number of unit cells of graphene within a unit cell of the superlattice, $N_A^H$ is the number of unit cells where the potential is higher (see Fig. 9), and $a$ is the lattice constant ($= 2.46$ Å). If intervalley coupling is ignored, the minimum value of $V_{SL}$ which allow new zero-energy points to emerge around the original Dirac point is [45]

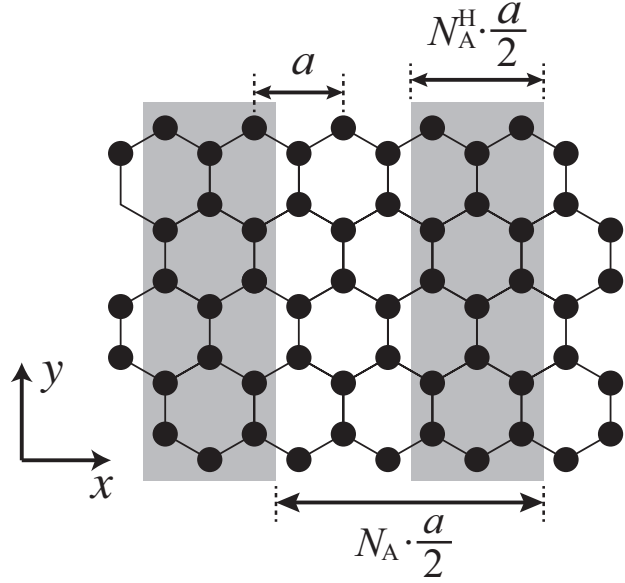

**Figure 9** A schematic of an armchair-type graphene superlattice. The on-site potential at carbon atoms inside the shaded regions is higher than that outside.

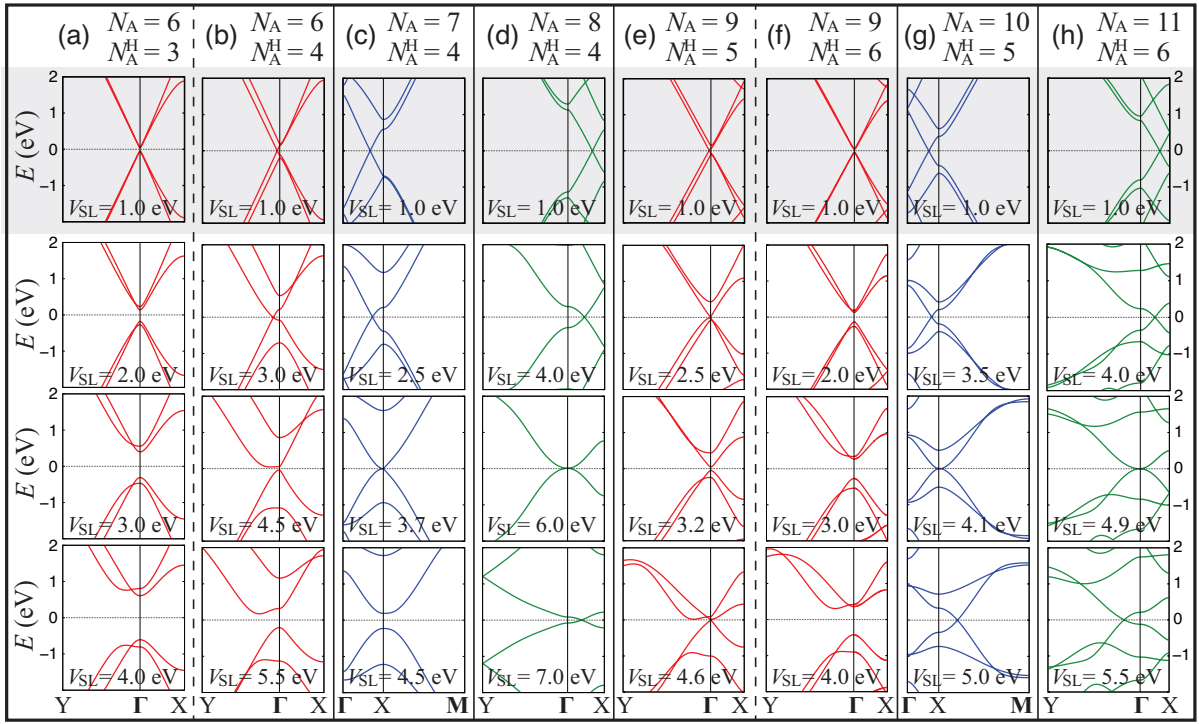

**Figure 10** The calculated energy band structures of armchair-type genuine graphene superlattices. $\Gamma = (0,0)$, $X = (2/N_A, 0)$, $Y = (0, 1/\sqrt{3})$, and $M = (2/N_A, 1/\sqrt{3})$ in units of $2\pi/a$. Dirac point energy is set to zero.

$$V_{\rm c} = \frac{4\pi\hbar v_{\rm F}}{N_{\rm A} \cdot a/2}, \tag{15}$$

where $v_{\rm F} = \sqrt{3}ta/2\hbar$ is the group velocity at the Dirac point of graphene without a superlattice potential (see Eq. (10)).

## 3.2. Armchair-type graphene superlattices

The electronic structure of genuine graphene superlattices is qualitatively similar to that of artificial graphene superlattices. If $N_{\rm A}$ is a multiple of three and so is $N_{\rm A}^{\rm H}$, the degeneracy at the Dirac point is lifted and a band gap opens up (Figs. 10(a), 10(f), and 11(b)). If $N_{\rm A}^{\rm H}$ is not a multiple of three, the Dirac point is split into two different zero-energy points moving along $k_y$ (Figs. 10(b), 10(e), and 11(c)). It is recently reported that a band gap opens up if $N_{\rm A}$ is a multiple of three and $N_{\rm A}^{\rm H}$ is exactly half of $N_{\rm A}$ [52]. According to our calculations, however, the necessary condition for the band gap opening up is that $N_{\rm A}$ and $N_{\rm A}^{\rm H}$ are both divisible by three (see, e.g., Fig. 10(f)).

We also observe the moving (along $k_x$, which is the periodic direction) and merging of the Dirac points in cases with $N_{\rm A} = 3n \pm 1$ (Figs. 10(c), 10(d), 10(g), and 10(h)). We focus on the superlattice with $N_{\rm A} = 10$ and $N_{\rm A}^{\rm H} = 5$ in the following. As $V_{\rm SL}$ increase, two Dirac points approach each other and finally merge at X (Figs. 12(a) and 12(b)); and then, it splits into two zero-energy points lying on X − M (Fig. 12(c)). The effective mass of the charge carriers at the Fermi energy either vanishes or remains finite depending on $V_{\rm SL}$. This evolution in the positions of the zero-energy points in momentum space (Fig. 12) may result in the change of transport properties of heterostructures containing graphene superlattices.

These characteristic features are not predicted in the theories adopting single-valley approximation [45, 46, 50] and also the similarity between these features of genuine and artificial graphene suggests that the contribution from the eigenstates with high energy is not significant; hence, they should be attributed to intervalley coupling. The effect of intervalley coupling is significant also in sinusoidal-type graphene superlattices whose on-site potential is

$$V_i = V_{\rm SL} \sin\left[\frac{4\pi}{N_{\rm A} a}\left(x_i - \frac{a}{4}\right)\right]. \tag{16}$$

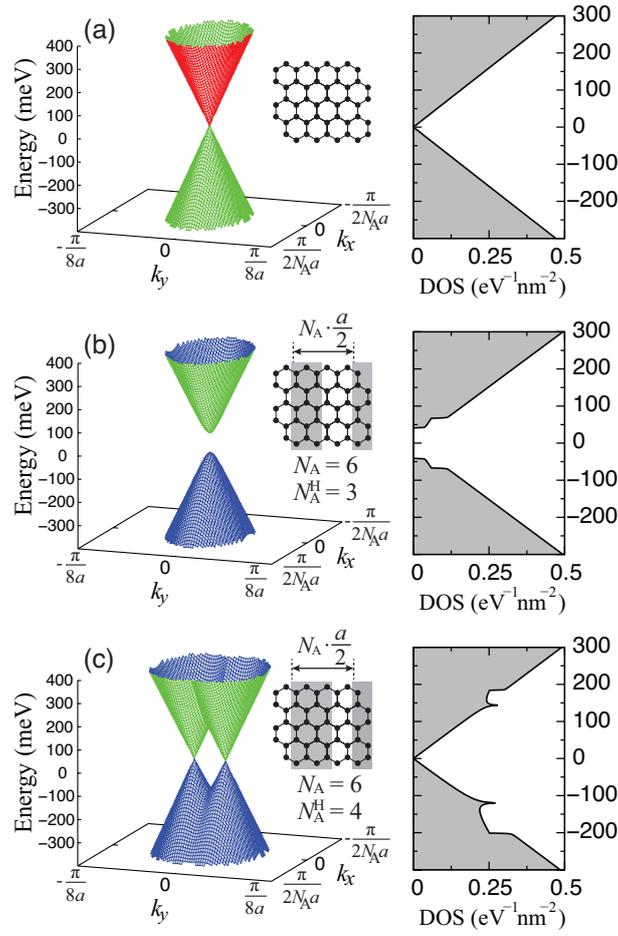

**Figure 11** (a) The energy band structure around Γ and density of states of armchair-type genuine graphene superlattice with $N_A = 6$ when no additional one-dimensional potential is applied. (b) and (c) Similar quantities as in (a) but with $N_A^H = 3$ and $N_A^H = 4$, respectively. We have used $V_{SL} = 1.0$ eV.

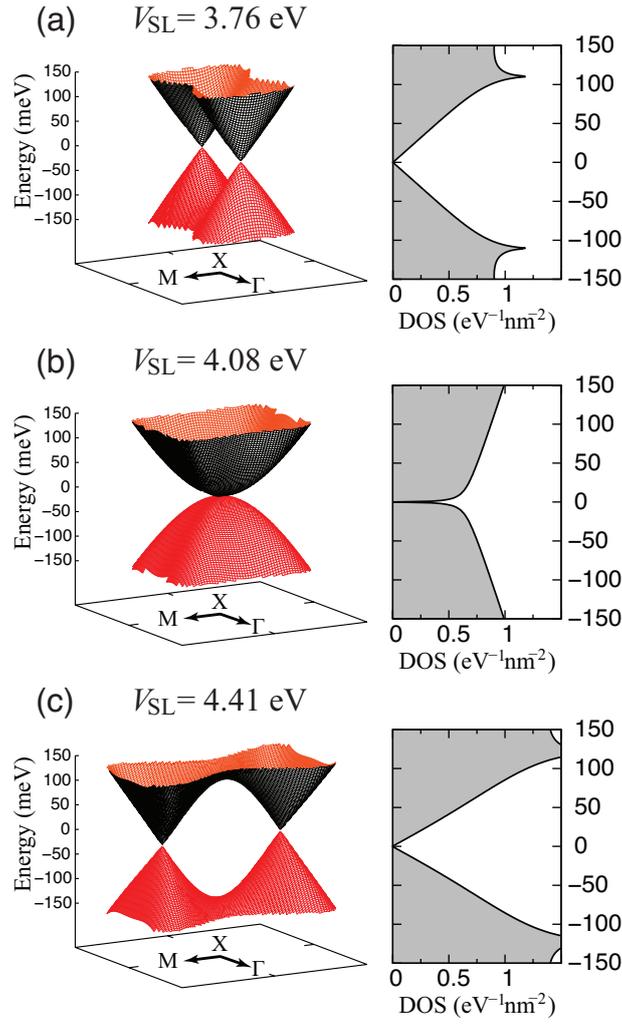

**Figure 12** The energy band structures around X and densities of states of armchair-type genuine graphene superlattices with $N_A = 10$ and $N_A^H = 5$.

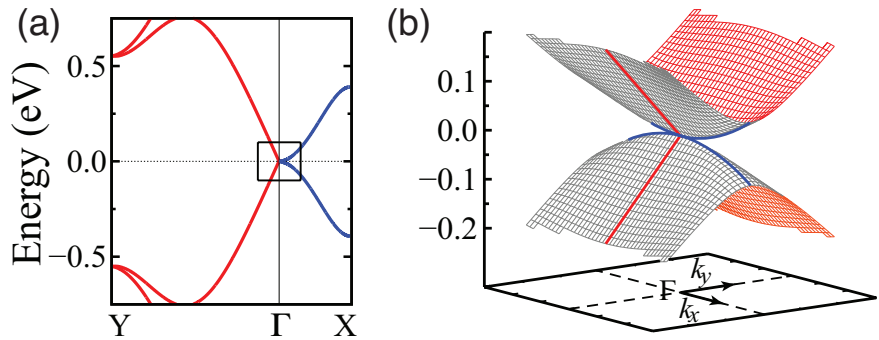

**Figure 13** The energy band structures around $\Gamma$ of graphene under a sinusoidal superlattice periodic potential along $x$ (Eq. (16)). We have used $N_A = 10$ and $V_{SL} = 5.0$ eV. The box inside (a) illustrates the energy-momentum range used in (b).

Figure 13 shows the electronic structure of such a superlattice, in which $v_x$, the group velocity along the periodic direction of the superlattice potential, vanishes while $v_y$ remains finite. This result is exactly the opposite of what is expected in the previous studies adopting single-valley approximation [46, 53]: finite $v_x$ and vanishing $v_y$. In the sections below, we confine our discussion to Kronig-Penneytype graphene superlattices (Eq. (14)).

### 3.3. Intervalley coupling

Let $\mathbf{k}_D$ be the Bloch wavevector of the Dirac point a graphene superlattice which is at **K** if $V_{SL} = 0$ ($\mathbf{k}_D \neq \mathbf{K}$ if $V_{SL} \neq 0$). An eigenstate of the superlattice at a Dirac point, $|\mathbf{k}_D, n\rangle$, can be written in terms of the eigenstates of graphene, $|\mathbf{k}_D, m\rangle_0$, where $m$ and $n$ are band indices:

$$|\mathbf{k}_D, n\rangle = \sum_{l,m} c_{n,l,m} |\mathbf{k}_D - l\mathbf{b}_1, m\rangle_0,$$

where $\mathbf{b}_1 = (4\pi/N_A a, 0)$ (Eq. (8)) and the summation runs over all integers $l$ and band indices $m$ ($= \pm 1$). We define the opposite valley contribution as $\sum_{n,l',m} |c_{n,l',m}|^2/2$, where the summation runs over integers $l'$ with which $\mathbf{k}_D - l'\mathbf{b}_1$ is inside the first Brillouin zone and is close to K′ than to K and band indices $n$ and $m$ ($\pm 1$) (Figs. 14(b), 14(c), and 14(d)). Only when the opposite valley contribution we defined is close to zero can intervalley coupling be safely ignored.

The opposite valley contribution increases with $V_{SL}$; if $V_{SL} \geq \tilde{V}_c$, the eigenstates of graphene belonging to two different valleys contribute equally (50 %) to the eigenstate of the superlattice (Figs. 14(b), 14(c), and 14(d)). This result suggests that intervalley coupling is essential for explaining the moving and merging of the Dirac points and that the valley index is not a good quantum number describing the eigenstates of the superlattices.

We find that the results are similar in cases with $N_A = 3n - 1$ (not shown).

### 3.4. Competition between $V_c$ and $\tilde{V}_c$

$V_c$ is the value of $V_{SL}$ at which the new zero-energy points are created within single-valley approximation (Eq. (15)), while $\tilde{V}_c$ is the value at which the two Dirac points merge either at Γ (if $N_A = 3n - 1$) or X (if $N_A = 3n + 1$). The dependences of $V_c$ and $\tilde{V}_c$ on the spatial period, or $N_A$, are qualitatively different from each other. According to Eq. (15), $V_c$ is inversely

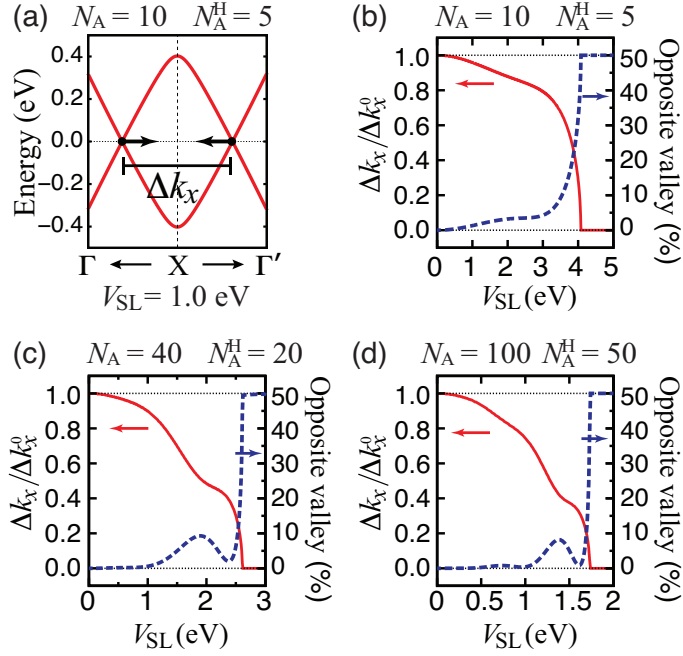

**Figure 14** (a) The energy band structure of an armchair-type genuine graphene superlattice around X. $\Gamma' = (8\pi/N_A a, 0)$. (b)-(d) The distance between two Dirac points in momentum space, $\Delta k_x$ illustrated in (a), divided by $\Delta k_x$ at $V_{SL} = 0$, $\Delta k_x^0$ (solid or red curve) and the opposite valley contribution (dashed or blue curve; see Sec. 3.3) versus $V_{SL}$ proportional to the spatial period. On the other hand, $\tilde{V}_c$ does not strongly depend on the spatial period. [The reason is as follows. First, the energy shift arising from the effects of intervalley coupling is inversely proportional to the spatial period, and the energy-momentum dispersion is linear around the Dirac point. Therefore, (i) the displacement of $\mathbf{k}_D$ by the energy shift is inversely proportional to the spatial period. Also, it is obvious that (ii) the width of the Brillouin zone along $k_x$, the periodic direction of the superlattice potential, is inversely proportional to the spatial period. From (i) and (ii), we conclude that $\tilde{V}_c$ is almost independent of the spatial period.] The results of the calculations in cases $N_A = 10$, 40, and 100 illustrate that $\tilde{V}_c$ indeed does not change much with $N_A$ (Figs. 14(b), 14(c), and 14(d)).

This difference in the dependencies on the spatial period brings about a competition between $V_c$ and $\tilde{V}_c$. If $V_c$ is lower than $\tilde{V}_c$, new zero-energy points are created around the original Dirac point at $V_{SL} \approx V_c$ (Figs. 15(a) and 15(c)). For instance, $V_c = 1.30$ eV and $\tilde{V}_c \approx 3.05$ eV in the case of the armchair-type genuine graphene superlattice with $N_A = 50$ and $N_A^H = 25$. Thus, the zero-energy points are created (Figs. 15(c) and 15(d)). If, on the other hand, $V_c$ is higher than $\tilde{V}_\mathrm{c}$, we cannot observe such a creation because the two Dirac points merge at $V_\mathrm{SL} = \tilde{V}_\mathrm{c}$ (Fig. 12(b)).

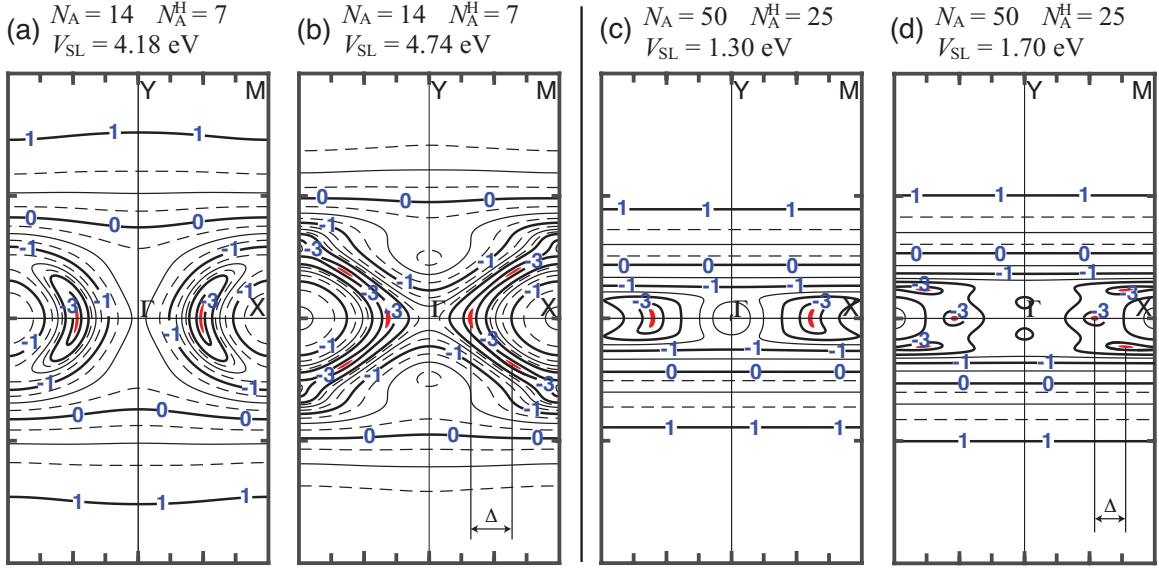

**Figure 15** $\mathrm{Log}_e(\Delta E/\mathrm{eV})$, where $\Delta E$ is the difference between the energy eigenvalues of the lowest-energy conduction band and the highest-energy valence band. The regions where the logarithm is lower than -4 are shown in red. The widths of the Brillouin zone in the $k_x$ direction are exaggerated for clarity.

In this context, we define the critical spatial period, $L_\mathrm{c}$, as the minimum spatial period with which such a creation of new zero-energy points can occur in the electronic structure of the superlattice. This quantity depends on the type of the superlattice; for an armchair-type superlattices, we found that such critical spatial period is about six times the lattice constant, $a$. $L_\mathrm{c} \sim 1.5$ nm in the case of a genuine graphene superlattice and $\sim 12$ nm in the case of an artificial graphene superlattice with $a = 2$ nm.

Intervalley coupling has a significant effect also on the newly created zero-energy points. (The effects of intervalley coupling on the original Dirac points were discussed in Sec. 3. 3) According to the previous studies adopting single-valley approximation which anticipated the emergence of new zero-energy points, such new zero-energy points move along the $k_y$ direction, which is perpendicular to the periodic direction of the superlattice potential [45, 46, 50] (also see Sec. 2.3). However, we find that the Bloch wavevector of the new zero-energy point changes not only in the $k_y$ direction, but also in the $k_x$ direction as shown in Figs. 15(b) and 15(d).

Furthermore, we verified that a similar change in $V_\mathrm{SL}$ brings about a similar change in $k_x$ of the newly created zero-energy point divided by the width of the Brillouin zone of the superlattice along $k_x$, regardless of the spatial period of the superlattice (see $\Delta$ in Figs. 15(b) and 15(d)). As the original Dirac points merge at Γ or X, the newly created zero-energy point merges with the other one approaching from the opposite side (not shown).

## 4. Conclusion

We have investigated the rich electronic structures of artificial and genuine graphene superlattices. The characteristic features including band-gap openings, substantial changes in the effective mass of the charge carriers, and displacements in momentum space of the Dirac cones of such superlattices are largely determined by the periodic direction, spatial period, and strength of the superlattice potentials.

We verified that the characteristic features we observed in the calculated electronic structures can be better explained if we take into account the effects of intervalley coupling. Furthermore, contrary to the assumptions made in the many previous studies, the changes in the electronic energy eigenvalues of graphene superlattices arising from the effects of intervalley coupling cannot be ignored no matter how long the spatial period of the superlattice is, because both the changes in energy eigenvalues and in energy bandwidths are inversely proportional to the spatial period.

Finally, we note that our theory is directly relevant to real experiments because even the artificial graphene superlattices can already be fabricated using currently available techniques [17-19].

**Acknowledgements**

This work was supported by Korean NRF-2013R1A1A1076141 funded by MSIP and computational resources were provided by Aspiring Researcher Program through Seoul National University in 2014.